\documentclass[twocolumn ]{aastex631}

\usepackage{amsmath}

\newcommand{\msun}{\ensuremath{\, \mathrm M_{\sun{}}}}

\newcommand{\new}[1]{ #1}
\received{July 6, 2021}
\revised{August 6, 2021}
\accepted{August 10, 2021}

\submitjournal{ApJL}

\shorttitle{Competition of gas and stars in MBHB evolution}
\shortauthors{Bortolas et al.}

\graphicspath{{./}{img/}}
\begin{document}

\title{The competing effect of gas and stars in massive black hole binaries evolution}

\correspondingauthor{Elisa Bortolas}
\email{elisa.bortolas@unimib.it}

\author[0000-0001-9458-821X]{Elisa Bortolas}
\affiliation{Dipartimento di Fisica ``G. Occhialini", Universit\'a degli Studi di Milano-Bicocca \\ Piazza della Scienza 3 \\ 20126 Milano, Italy}
\affiliation{INFN, Sezione di Milano-Bicocca \\ Piazza della Scienza 3 \\ 20126 Milano, Italy}

\author[0000-0002-8400-0969]{Alessia Franchini}
\affiliation{Dipartimento di Fisica ``G. Occhialini", Universit\'a degli Studi di Milano-Bicocca \\ Piazza della Scienza 3 \\ 20126 Milano, Italy}
\affiliation{INFN, Sezione di Milano-Bicocca \\ Piazza della Scienza 3 \\ 20126 Milano, Italy}

\author[0000-0001-7889-6810]{Matteo Bonetti}
\affiliation{Dipartimento di Fisica ``G. Occhialini", Universit\'a degli Studi di Milano-Bicocca \\ Piazza della Scienza 3 \\ 20126 Milano, Italy}
\affiliation{INFN, Sezione di Milano-Bicocca \\ Piazza della Scienza 3 \\ 20126 Milano, Italy}

\author[0000-0003-4961-1606]{Alberto Sesana}
\affiliation{Dipartimento di Fisica ``G. Occhialini", Universit\'a degli Studi di Milano-Bicocca \\ Piazza della Scienza 3 \\ 20126 Milano, Italy}

\begin{abstract}

Massive black hole binaries are predicted to form during the hierarchical assembly of  cosmic structures and will represent the loudest sources of low-frequency gravitational waves (GWs) detectable by present and forthcoming GW experiments.
Before entering the GW-driven regime, their evolution is driven by the interaction with the surrounding stars and gas. While stellar interactions are found to always shrink the binary, recent studies predict the possibility of binary outspiral mediated by the presence of a gaseous disk, which could endlessly delay the coalescence and impact the merger rates of massive binaries.
Here we implement a semi-analytical treatment that follows the binary evolution under the combined effect of stars and gas. We find that binaries may outspiral only if they accrete near or above their Eddington limit and only until their separation reaches the gaseous disk self-gravitating radius. 
Even in case of an outspiral, the binary eventually reaches a large enough mass for GW to take over and drive it to coalescence. 
The combined action of stellar hardening, mass growth and GW-driven inspiral brings binaries to coalescence in few hundreds Myr at most, implying that gas-driven expansion will not severely affect the detection prospects of upcoming GW facilities. 

\end{abstract}

\keywords{Supermassive black holes (1663), Stellar dynamics (1596), Gravitational wave sources (677), Galaxy accretion disks (562),   Computational astronomy (293)}

\section{Introduction}\label{sec:intro}

In the past two decades it has been established that massive black holes (MBHs) inhabit the center of massive galaxies \citep{Kormendy2013}. As galaxies grow by merging with other galaxies along the cosmic history, MBH binaries (MBHBs) are considered to be a natural outcome of the hierarchical structure formation scenario. In fact, following a galaxy merger, the two MBHs hosted by the parent galaxies find themselves inside the same host galaxy \citep{Begelman1980}.

The  evolution of such  MBH pairs critically depends on the characteristics of the surrounding environment: after a successful dynamical friction-driven inspiral \citep{Chandrasekhar1943, Capelo_Dotti_2017, Pfister2019, Bortolas2020, Bonetti2020, Bonetti2021},  bound binaries form in the central parsecs of galaxies. MBHBs can then further shrink down to the gravitational wave (GW) emission stage  through different mechanisms. The most explored ones involve the interaction with either the stellar background \citep[e.g.][]{Khan2011,  Vasiliev2014,Bortolas2016,Gualandris2017, Bortolas2018, Bortolas2018tr, Varisco2021} or a gaseous circumbinary disk 
\citep[e.g.][]{Armitage2002,Escala2004,Dotti2006, Mayer2007,Lodato2009, Cuadra2009}. 
Generally, the evolution is studied either in purely stellar or purely gaseous environments (with the notable exception of \citealt{Kelley2017b, Kelley2019}), but \new{in reality, often} both components are present at the same time.
While  the evolution driven by stars is always found to shrink the binary via three-body stellar encounters \citep[e.g.][]{Quinlan1996}, the effect of the interaction   with a gaseous disk is still under debate.  
The picture outlined in early simulations \citep[e.g.][]{artymowicz1994}, that coherently points towards binary shrinking in response to the interaction with a gaseous disk, has been recently challenged by several studies \new{\citep{Munoz2019,Duffell2020, Moody2019}}. These recent hydrodynamical simulations find possible binary expansion within the range of binary and disk parameters explored. 

More recently, it has been pointed out by \citet{Tiede2020} and then confirmed with 3D hydrodynamical simulations of a live binary by \citet{Heath2020} that the evolution of the binary semi-major axis depends on the temperature of the disk. Binaries tend to expand only for relatively thick disks, with aspect ratio $H/R\gtrsim 0.05-0.2$, the exact threshold depending on the details of the numerical approach employed.

Irrespective of the specific disk configurations, a possible ``outspiral'' phase is critically relevant in view of the present and future low-frequency GW observatories that are expected to unveil coalescing binaries across the Universe, as  Pulsar Timing Arrays (PTAs; \citealt{2016MNRAS.458.3341D,2016MNRAS.455.1751R,2019MNRAS.490.4666P,2021ApJS..252....5A}), and the Laser Interferometer Space Antenna (LISA; \citealt{AmaroSeoane2017,Schodel2017LISA,Barack_et_al_2019}). Stalled or expanding binaries \new{could} dramatically impact the expected coalescence rates for the forthcoming facilities, therefore affecting the detection prospects and ultimately their mission science cases. On the other hand, long-lived outspiralling binaries might be fairly frequent in the Universe, increasing the chance of detection in electromagnetic surveys.

In this paper, we combine the gas induced expansion found in the aforementioned studies \citep{Munoz2019,Munoz2020,Duffell2020} with the stellar and GW hardening, to verify in which regime the binary is found to possibly stall or expand its separation under the concurrent effect of different evolutionary processes. We consider circular, equal mass binaries and an infinite supply of gas \new{(see also \citealt{Munoz2020}, who compared the orbital decay of MBHBs interacting with finite or infinite supplies of gas)}, which are idealised assumptions aimed at maximizing the effect of gas outspiral. Nevertheless, we anticipate that we found the bound binary orbit to always decay in a relatively short timescale ($\sim 10^8 $ yr), indicating that the gas-induced binary expansion can only mildly delay the final coalescence, and that gas-driven expansion necessarily has to revert into a shrinking as the binary mass gets large enough due to gas accretion.

The paper is organised as follows: in Sec.~\ref{sec:methods} we describe the semi-analytical approach used to study the binary semi-major axis evolution; in Sec.~\ref{sec:results} we outline and comment the obtained results; in Sec.~\ref{sec:discussion} we discuss the possible implications of our findings together with the possible caveats and we  draw our conclusions.

%%%%%%%%%%%%%%%%%%%%%%%%%%%%%%%%%%%%%%%%%%%%%%%%%%
%%%%%%%%%%%%%%%%%%%%%%%%%%%%%%%%%%%%%%%%%%%%%%%%%%

\section{Theoretical framework}\label{sec:methods}

In this section, we present the different prescriptions  describing the binary semi-major axis evolution, $\dot{a}$, as a result of interactions with stars, gas, and of GW emission.

\subsection{Binary evolution in a stellar environment}\label{sec:stellar_hardening}

The stellar driven binary hardening rate can be written as \citep{Quinlan1996,Sesana2015}
\begin{equation}
    \dot{a}_{\star} = -\frac{H G\rho}{\sigma}\,a^2
    \label{eq:stellar_hardening}
\end{equation}
where $\rho$ and $\sigma$ are the stellar density and velocity dispersion at the binary influence radius respectively, $H$ is a dimensionless quantity that mildly depends on the ratio $a/a_{\rm h}$ and on the binary mass ratio $q=m_2/m_1\leq1$, where $m_1$, $m_2$ are the masses of the two binary components; we also set $m=m_1+m_2$. The specific value of $H$ ranges between 12 and 20 for binary separations $a<a_{\rm h}$, where $a_{\rm h} = G m_2/(4 \sigma^2)\propto m^{0.543}$ (see below)
is the so-called hard binary separation, i.e. the separation below which the hardening in purely stellar environments is dominated by three body  interactions, and is well described by Eq.\ref{eq:stellar_hardening}. This constant is generally computed using scattering experiments, in particular we here use the fit obtained by \citet{Sesana2006}.

In order to obtain a general description of the stellar hardening, we need to express $\rho$ and $\sigma$ in terms of the binary mass $m$. We make the conservative assumption that no nuclear stellar cluster is present in the system, so that stellar hardening is solely due to interactions of the binary with the bulge structure of the galaxy. We thus express the  $\sigma$ as \citep{Kormendy2013}
\begin{equation}
    \sigma = 200 \left( \frac{m}{3.09 \times 10^8\msun }\right)^{\frac{1}{4.38}} {\rm km\ s}^{-1},
\end{equation}
while we write the central density as  $\rho=2m/(4/3\pi r_{\rm infl}^3)\propto m^{-0.68} $, where $r_{\rm infl}$ is the influence radius containing twice the binary mass, that can be obtained from scaling relations and is equal to 
$    r_{\rm infl}= 35 \left( \tfrac{m}{10^8\msun }\right)^{0.56} {\rm pc}$ \citep{Merritt2009}. This implementation implies $\dot{a}_\star\propto m^{-0.91}$, so that the hardening is less efficient for more massive binaries.

\subsection{Binary evolution in a gaseous environment}\label{sec:gas_hardening}

According to the 2D hydrodynamical simulations performed by \citet{Munoz2019,Munoz2020}, the binary semi-major axis evolution can be written as
\begin{equation}
    \dot{a}_{\rm gas} =  2.68 \frac{\dot{m}}{m} a
    \label{eq:gas_hardening}
\end{equation}
where $\dot{m}$ is the binary mass accretion rate.\footnote{\new{We note that Eq.~(\ref{eq:gas_hardening}) is valid in a regime where the change in accretion rate occurs over a longer timescale compared to the viscous time.  It is easy to verify that this condition is generally fulfilled in cold accretion disks.}} This is essentially eq. 9 in \citet{Munoz2020}\footnote{See the upper panel of their fig.~7.} and is formally valid only for $q\gtrsim0.1$ (\new{\citealt{Duffell2020} explore gas-driven evolution for $q=0.01-1$ and also find positive, nearly constant torques for $q\gtrsim 0.1$}) for a prograde binary located in the disk plane.

In our controlled experiment, we are free to choose the mass accretion rate $\dot{m}$. We explore two distinct possibilities:
\begin{enumerate}
    \item a fixed physical value $\dot{m}$ throughout the binary evolution, so that the mass flux onto the binary is the same even if the binary mass increases
    \item a fixed fraction $f_{\rm Edd}$ (Eddington ratio) of the Eddington 
    accretion rate, $\dot{m}=f_{\rm Edd} \dot{m}_{\rm Edd}$, where
    \begin{equation}
    \begin{split}
    \label{eq:mdotedd}
        \dot{m}_{\rm Edd} = \frac{4\pi G m\, m_{\rm p} }{\eta \sigma_{\rm T} c} = 2.26\times10^{-2} \left(\frac{ \eta }{0.1}\right)^{-1}\\
        \left(\frac{m}{10^6 \msun}\right) \msun\,\rm yr^{-1}.
\end{split}        
    \end{equation}
    Here $m_{\rm p}$ is the proton mass, $c$ is the speed of light, $\sigma_{\rm T}$ is the Thomson scattering cross section and $\eta$ is the accretion efficiency, which we assume to be 10\%. Note that in this case the accretion rate onto the binary does increase with the binary mass.
\end{enumerate}

If the binary components do not have the same mass, the lighter component is typically found, in numerical simulations, to accrete mass more efficiently compared to the heavier component. The consequence is that all the binaries tend to evolve towards equal mass. In this paper we thus only show cases for which $q=1$ from the start. We tested the evolution accounting for the change in the binary mass ratio $\dot{q}$ for $q>0.1$, using the prescriptions inferred by \citet{Farris2014, Duffell2020}, and found no significant difference from the results presented here,   
as the binary becomes equal mass very quickly, in the gas dominated regime, before entering the GW-driven inspiral, which is the only process whose timescale is sensitive to $q$.

\subsection{Discs in the self-gravitating regime}

Circumbinary disks surrounding very massive black hole binaries, i.e. $m\gtrsim 10^7\msun$, are likely to be self-gravitating \citep{franchini2021}. When the disk self-gravity contribution is non negligible, the circumbinary disk can either self-regulate on a quasi-stable state through the formation of spirals or can fragment into clumps if the disk cooling mechanism is very efficient compared to the heating generated by the shocks induced by the spirals.
Since the   evolution in Eq.~\ref{eq:gas_hardening}  has been inferred for non-self-gravitating disks, 
we assume the binary   gas driven expansion (Eq.~\ref{eq:gas_hardening})  to be suppressed as the binary semi-major axis reaches the radius beyond which the disk self-gravity cannot be neglected.  This is motivated by the results of \citet{franchini2021}, who finds that, regardless of initial temperature, self-gravitating disk regulates themselves to $H/R<0.1$, promoting binary shrinking. The self-gravitating radius can be expressed as
\citep{Perego2009}
\begin{equation}
\begin{split}
    R_{\rm sg} &= 1.158\times10^{-2}f_{\rm Edd}^{-22/45} \left(\frac{\alpha}{0.1}\right)^{28/45} \left(\frac{\eta}{0.1}\right)^{22/45} \\ 
    & \left(\frac{m}{10^6\msun}\right)^{-7/45}\, \rm pc
\label{eq:Rsg}
\end{split}
\end{equation}
for a fixed the Eddington ratio, or as
\begin{equation}
\begin{split}
    R_{\rm sg} &= 1.817\times10^{-3}\,\left(\frac{\dot{m}}{1\msun/{\rm yr}}\right)^{-22/45} \left(\frac{\alpha}{0.1}\right)^{28/45}  \\
    & \left(\frac{m}{10^6\msun}\right)^{15/45}\, \rm pc
\label{eq:Rsg2}
\end{split}
\end{equation}
for a fixed accretion rate, where $\alpha$ is the disk viscosity parameter, which encapsulates the angular momentum transport mechanism within the disk \citep{SS1973}.

We assume the expansion to be suppressed  when the binary semi-major axis is above $C\, R_{\rm sg}$; here we set  $C=0.5$ since the radius of the disk cavity extends to $2a$, and therefore we neglect, as a first approximation, gaseous interaction once the inner disk cavity radius equals the disk self-gravitating radius.
We also tried larger values of $C$ finding no significant differences in the global evolution.

\new{The prescription for binary expansion may also break down at the radius beyond which the disc temperature drops below  $10^4$ K, possibly triggering the so-called ionization instability \citep[e.g.][]{Haiman2009}. This can occur at scales smaller than the self-gravitating radius. We however do not consider the ionization instability radius as a limit for the expansion as there is so far no evidence that gas-induced expansion would be hampered beyond this scale. This also means our treatment is rather conservative, as the binary expansion could in principle be prevented in a wider region of the parameter space.}

\subsection{Gravitational waves driven orbital decay}

If the binary semi-major axis drops below a certain threshold, then the subsequent evolution of the binary  is driven by GW emission.
Considering a zero orbital eccentricity, the binary semi-major axis  shrinks as \citep{Peters1964}
\begin{equation} \label{eq:gw_hardening}
     \dot{a}_{\rm GW} = - \frac{64}{5}\frac{G^3}{c^5}\,\frac{q}{(1+q)^2}\,\frac{m^3}{a^{3}}
\end{equation}
until the binary reaches coalescence.\\

\begin{figure}
    \centering
    \includegraphics[width=0.99\columnwidth]{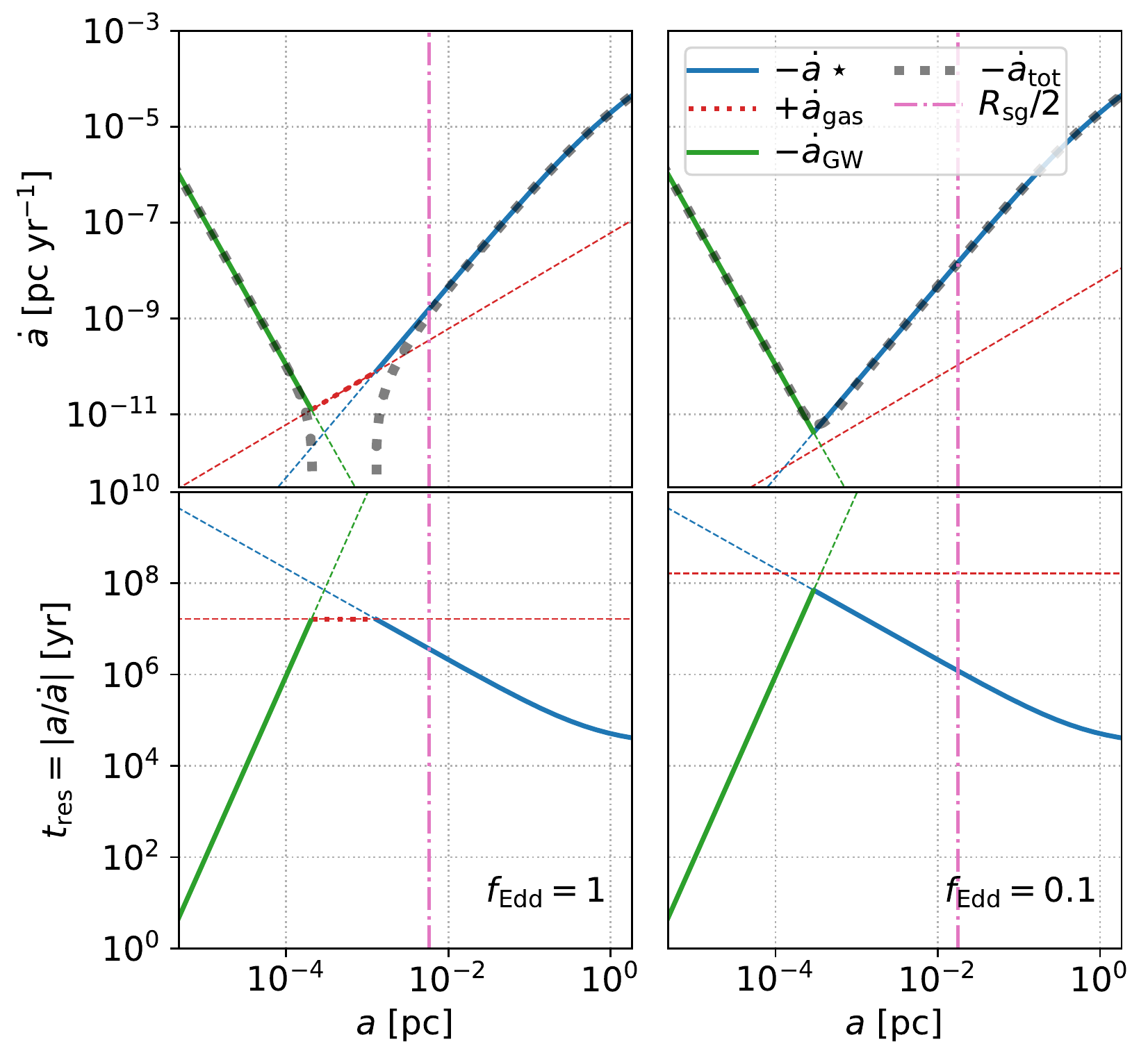}
    \caption{Hardening of a $10^6\msun$, equal mass binary modelled via the prescriptions detailed in Sec.~\ref{sec:methods}. The top panels show the binary shrinking (or expanding, in the case of gas) rate due to gas, stars and GW, and the cumulative hardening $\dot{a}_{\rm tot}$ (only shown when $<0$). The bottom panels show the associated residence timescale for all the components. Note that for the gas, the residence time has little meaning as gas tends to expand instead of shrinking the binary in the presented picture. We also show the self-gravitating radius of the disk (Eq.~\ref{eq:Rsg}) above which the gaseous hardening presented here does not hold anymore. The left panels show the evolution for a binary accreting at $\dot{m}_{\rm Edd}$ (whose shrinking is thus hindered by gas), the right-hand ones refer to a binary with  $\dot{m}=0.1\dot{m}_{\rm Edd}$, for which gas plays virtually no role. Very importantly,\textit{ the evolution shown here does not take into account the time variation of the binary mass}, which significantly changes the picture.}
    \label{fig:me6q1}
\end{figure}

Two examples of the hardening (or expansion) rate as a function of $a$ are presented in Fig.~\ref{fig:me6q1} for an equal mass, $10^6 \msun$ binary with different accretion rates. \new{The plots are produced by keeping the total binary mass $m$ fixed (although the employed prescription for $\dot{a}_{\rm gas}$ in Eq.~\ref{eq:gas_hardening} also accounts for the angular momentum change induced by mass accretion into the binary, \citealt{Munoz2020})}, so that the binary with $f_{\rm Edd}=1$ appears to stall once gas starts to dominate its evolution.

\subsection{Transition radii}

If we neglect for a moment the fact that the mass of the binary is not fixed in time, the different power-law dependencies of the binary hardening rates with $a$ imply that at large scales stellar hardening ($\dot{a}_\star\propto -a^{2}$) is always the dominant evolutionary mechanism. Conversely, at the smallest scales the binary is driven by GWs as $\dot{a}_{\rm GW}\propto -a^{-3}$. Gas driven evolution ($\dot{a}_{\rm gas}\propto a$), which may hinder the shrinking, can thus only be dominant between the stellar and GW driven evolution; depending on the involved parameters, it could also be always subdominant.

If gas is ever dominant, the transition between the star and gas dominated regimes occurs at 
\begin{align}
    a_{\rm \star \ \rightarrow \ gas} &=2.68\frac{\dot{m}}{m} \frac{\sigma}{HG\rho}.
\end{align}
If instead gas is always subdominant,  the GW driven hardening rate is equal to the stellar hardening rate at
\begin{align}
    a_{\rm \star \ \rightarrow \ GW} &=\left( \frac{64}{5}\frac{G^2}{c^5}\, \frac{\sigma}{H\rho}\frac{q}{(1+q)^2}\,m^3\right)^{1/5}.
\end{align}
Analogously, we can write the transition between the gas driven to the GW driven hardening to occur at
\begin{align}
 &a_{\rm gas \ \rightarrow \ GW} =1.48 \left(\frac{G^3}{c^5} \frac{q}{(1+q)^2} \frac{m^4}{\dot{m}}\right)^{1/4}.
\end{align}

\subsection{Threshold for gas-driven evolution} 

\begin{figure}
    \centering
    \includegraphics[width=0.95\columnwidth]{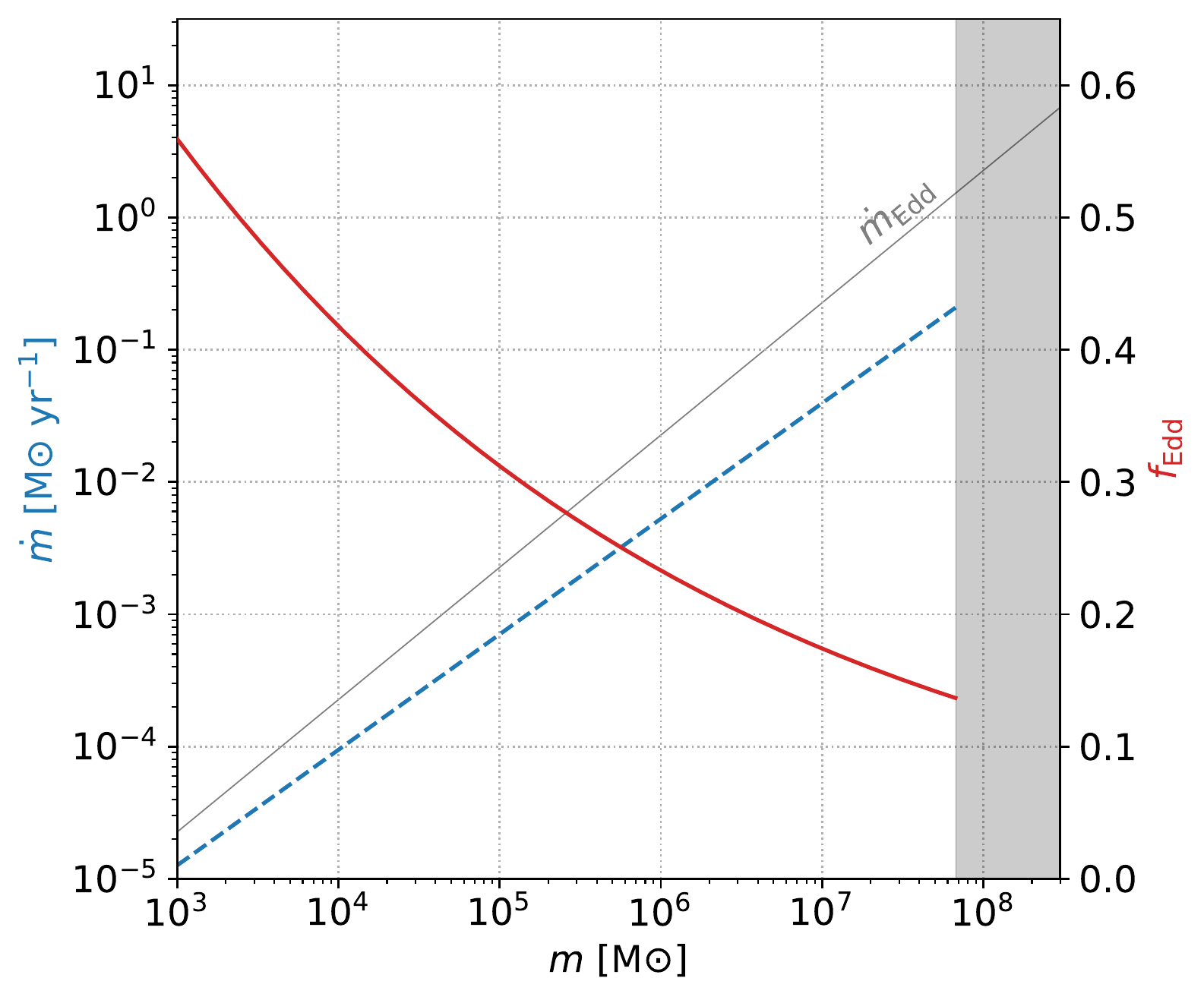}
    \caption{Minimum mass accretion rate above which gas-driven evolution becomes non negligible and the binary expansion can occur as a function of the total binary mass $m$. The blue dashed curve refers to the left-hand vertical axis and considers a fixed binary mass accretion rate (Eq.~\ref{eq:gaslim_mdot}). For reference, we also show the Eddington accretion rate $\dot{m}_{\rm Edd}$ with a solid grey line. The red line refers to the right hand vertical axis and expresses the accretion in terms of the Eddington ratio  (Eq.~\ref{eq:gaslim_fedd}).
    The curves are limited  by a grey vertical shaded region (Eq.~\ref{eq:gaslim_rsg}), which defines the masses above which the binary evolution can no longer be dominated by a non self-gravitating disk. All curves assume $q=1$.
    }
    \label{fig:gaslim}
\end{figure}

In the presented theoretical framework, ignoring for now the mass  change of the binary in time, we can define the region of the parameter space in which the gas-driven evolution is important. In particular, gas can play a non-negligible role if, at $a_{\rm \star \ \rightarrow \ GW}$, $|\dot{a}_{\rm gas}|> |\dot{a}_{\rm \star}| = |\dot{a}_{\rm GW}|$. This condition translates into:
\begin{equation}\label{eq:gaslim}
    \frac{\dot{m}}{m} > 0.621 \frac{G^{3/5}}{c} \left(\frac{q}{(1+q)^2}\right)^{1/5} \left(\frac{HG\rho}{\sigma}\right)^{4/5} m^{3/5}
\end{equation}
If we assume a fixed gas accretion rate $\dot{m}$, the gas dominates for accretion rates
\begin{equation}\label{eq:gaslim_mdot}
    \frac{\dot{m}}{\rm \msun\, yr^{-1}} > 5.41\times10^{-3} \left(\frac{H}{15}\right)^{4/5}
    \left(\frac{4q}{(1+q)^2}\right)^{1/5} \left(\frac{m}{10^6\msun}\right)^{0.87}.
\end{equation}
Instead, if we express the mass accretion rate as a function of the Eddington accretion rate $\dot{m}_{\rm Edd}$, we obtain a lower limit  for the Eddington ratio:
\begin{equation}\label{eq:gaslim_fedd}
    f_{\rm Edd}>0.24 \left(\frac{H}{15}\right)^{4/5}
    \left(\frac{4q}{(1+q)^2}\right)^{1/5} \left(\frac{m}{10^6\msun}\right)^{-0.13}.
\end{equation}

Note however that this only applies as long as the disk self-gravity can be neglected.
This translates into the additional condition that the self gravitating radius should be no smaller than the scale of the transition $a_{\star \ \rightarrow  \ \rm GW}$, which for the limiting values of accretion in Eqs. (\ref{eq:gaslim_mdot}, \ref{eq:gaslim_fedd}) implies:
\begin{equation}\label{eq:gaslim_rsg}
    m < 6.72\times 10^7 \msun \left(\frac{C}{0.5}\right)^{1.14} \left(\frac{H}{15}\right)^{-0.22}\left(\frac{4q}{(1+q)^2}\right)^{-0.34}.
\end{equation}
Above this critical mass, the disk self-gravity is important at all the scales where the gaseous disk induces binary evolution, so the modelling presented in Eq.\eqref{eq:gas_hardening} no longer holds and in the present model we just assume the gas to stop playing a role.  The boundaries of the parameter space relevant for gas driven evolution are shown in Fig. \ref{fig:gaslim}.

%%%%%%%%%%%%%%%%%%%%%%%%%%%%%%%%%%%%%%%%%%%%%%%%%%
%%%%%%%%%%%%%%%%%%%%%%%%%%%%%%%%%%%%%%%%%%%%%%%%%%

\subsection{Differential equations for the binary evolution in time}

Below we present the time evolution of the binary semi-major axis induced by stars, gas and GWs. In particular, we solve a set of differential equations for $\dot{a} = \dot{a}_\star + \dot{a}_{\rm gas}+ \dot{a}_{\rm GW}$ (Eq.~\ref{eq:stellar_hardening},~\ref{eq:gas_hardening},~\ref{eq:gw_hardening}) and $\dot{m}$. The mass accretion rate would be either expressed as a fixed value in time, or by fixing the Eddington ratio $f_{\rm Edd}$.  
In order to avoid numerical issues, we smooth out the binary gas driven evolution at $CR_{\rm sg}$:
\begin{equation}
    \dot{a}_{\rm gas, step} =\dot{a}_{\rm gas}\left(1+\tanh[A(1-a/(CR_{\rm sg}))]\right)
\end{equation}
where $A$ is a constant that we set equal to 20. In the present framework, we always assume that the binary can interact with an infinite gas reservoir, so that we maximise the possible effects of gas-driven expansion.
We start our integration at $a_0=10a_h$ (Sec.~\ref{sec:stellar_hardening}) and we stop it when the binary is well within the GW dominated phase. Here we also neglect the binary eccentricity evolution, and we assume an always circular binary. 

\section{Results}\label{sec:results}

\begin{figure}
    \centering
    \includegraphics[width=1\columnwidth]{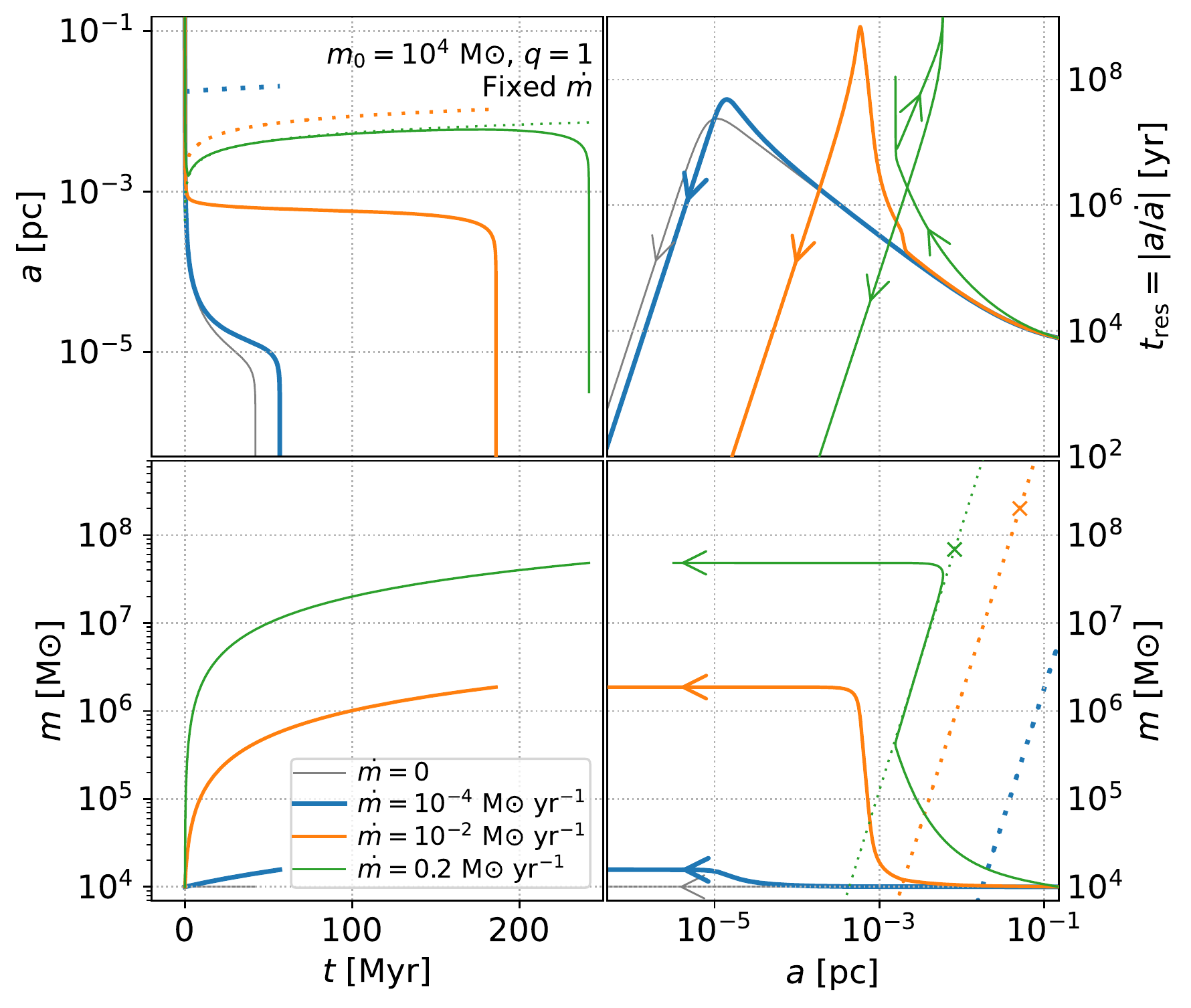}
    \caption{Evolution of a  binary  resulting from the interaction with gas, stars and from the emission of GWs. The binary has an initial mass of $10^4\msun$, and $q=1$ throughout the evolution. The plots here show three cases in which the binary is assumed to have a fixed mass accretion rate, whose values are shown in the legend. The left-hand panels show the time evolution of the binary semi-major axis (top) and total mass (bottom). The right hand panels show, as a function of the binary semi-major axis, the modulus of \new{ the residence time $|a/\dot{a}|$} (top) and the binary mass (bottom). In the top-left and bottom-right panel, the dotted lines show the  self-gravitating radius as a function of time and mass, respectively, for the assumed mass accretion rates. In the right-hand panels, the arrows show the direction of the evolution; in particular, in the $|a/\dot{a}|$ versus $a$ plot, while cases with $\dot{m}<0.2\msun$ yr$^{-1}$ should be read from right to left as the $\dot{a}$ is always negative and the binary never expands its separation, in the case with $\dot{m}=0.2\msun$ yr$^{-1}$ the binary initially shrinks, then it undergoes a phase of expansion, and finally  it starts shrinking again. The crosses in the bottom-right panel mark the position of $m_{\rm max}$ (Eq.~\ref{eq:mmax_fixmdot}).
    }
    \label{fig:evol_fixed_mdot}
\end{figure}

\begin{figure}
    \centering
    \includegraphics[width=1\columnwidth]{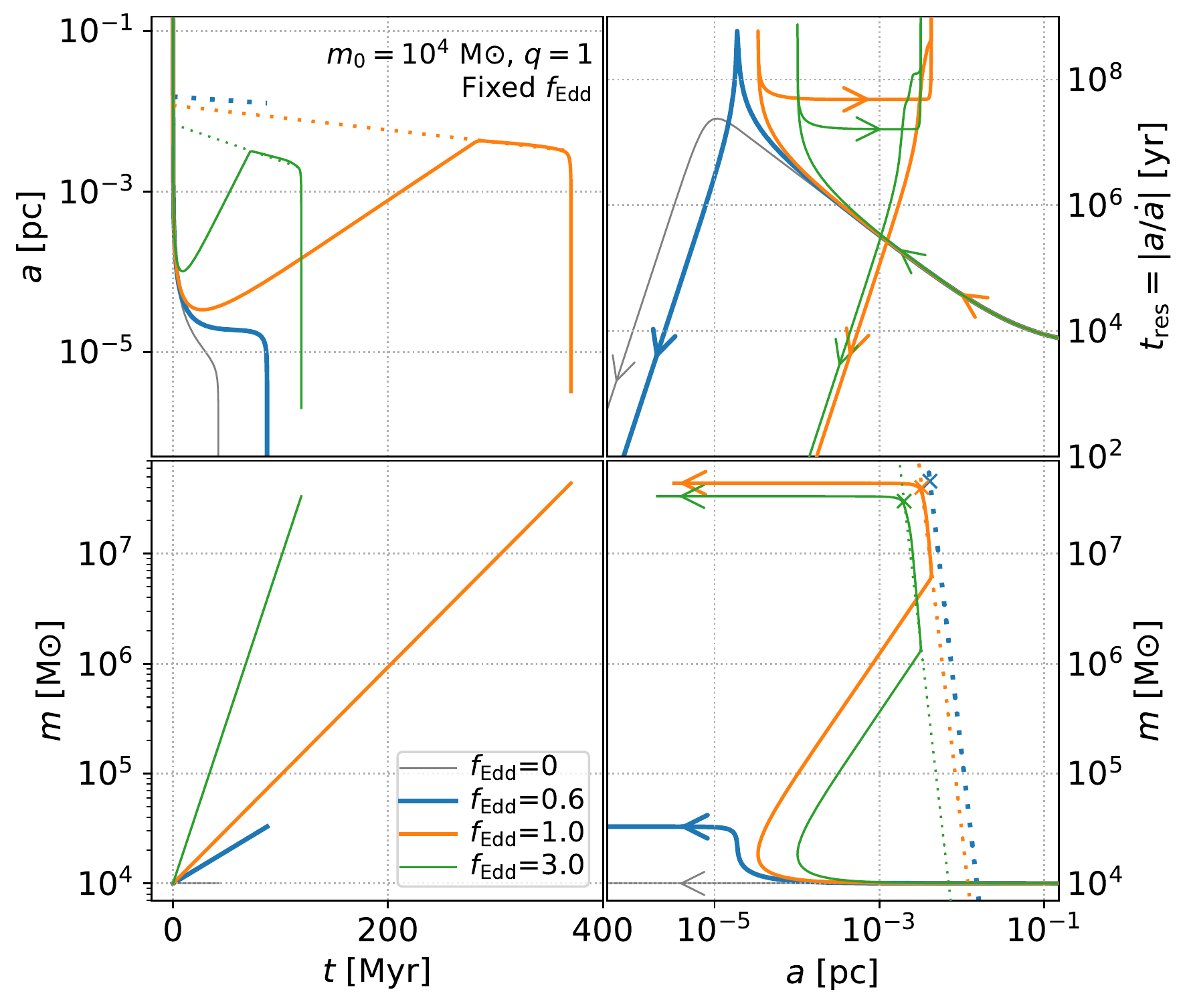}
    \caption{Same as Fig.~\ref{fig:evol_fixed_mdot} for a binary whose Eddington ratio is fixed along the evolution, as detailed in the legend. The  $x$s in the bottom right panel refer to the mass limit in Eq.~\ref{eq:mmax_fixf}.}
    \label{fig:evol_fixed_fedd}
\end{figure}

\begin{figure}
    \centering
    \includegraphics[width=1\columnwidth]{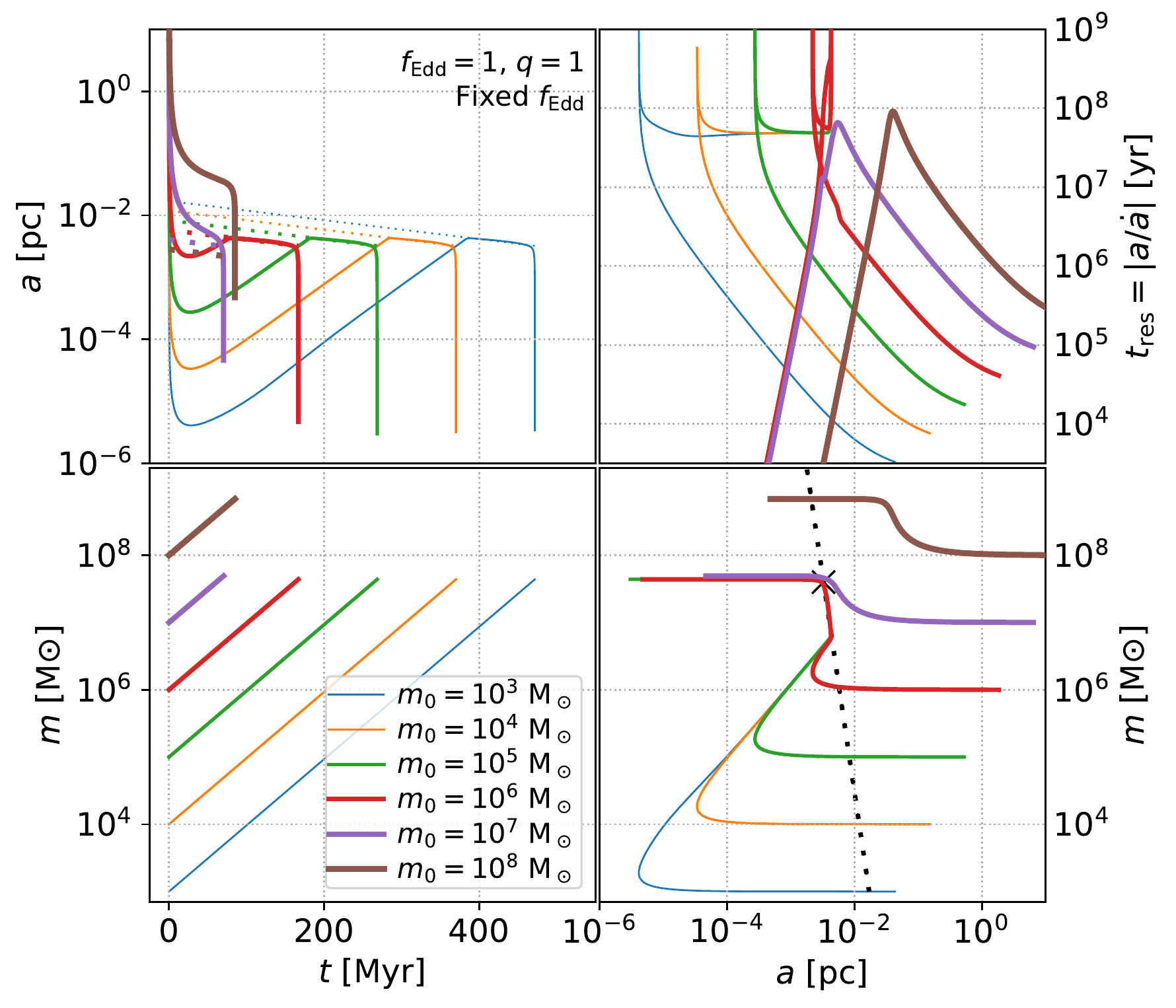}
    \caption{\new{Same as Fig.~\ref{fig:evol_fixed_mdot} for a binary accreting always at the  Eddington limit (fixed $f_{\rm Edd}=1$), varying its initial mass, as detailed in the legend. The $\times$ in the bottom right panel refers to the mass limit in Eq.~\ref{eq:mmax_fixf}.}}
    \label{fig:evol_fixed_m}
\end{figure}

Figures~\ref{fig:evol_fixed_mdot}~and~\ref{fig:evol_fixed_fedd} show different aspects of the binary evolution for an equal mass binary with initial mass of $10^4\msun$ and different mass accretion rates\footnote{The chosen initial mass is relatively small, but  a lower initial mass makes the effects of gas-driven expansion  more extreme for a given $f_{\rm Edd}$ or $\dot{m}$, as the binary can grow for a longer time before its mass gets large enough so that the surrounding disk becomes self-gravitating \new{(see e.g. Fig.~\ref{fig:evol_fixed_m}).} }. In particular, Fig.~\ref{fig:evol_fixed_mdot} assumes the binary to have a fixed $\dot{m}$ along the evolution, while Fig.~\ref{fig:evol_fixed_fedd} assumes a fixed Eddington ratio. The plots show that, if the mass accretion rate is small enough, the binary semi-major axis decreases as a result of stellar and then GW hardening, without the gas playing a significant role.

The impact of gas becomes stronger as the binary accretes mass more efficiently (i.e. $\dot{m}$ or $f_{\rm Edd}$ get larger). While the initial evolution is always driven by stellar interactions that shrink the binary ($\dot{a}<0$), the gas driven expansion may counterbalance the stellar 
hardening so that the binary shrinking turns into an expansion, delaying GW-driven hardening (green line in Fig.~ \ref{fig:evol_fixed_mdot}, green and orange line in Fig.~\ref{fig:evol_fixed_fedd}).

Note that the reason for the expansion (rather than a stalling) is different in the two accretion scenarios. When the binary accretes at a fixed $f_{\rm Edd}$,  $\dot{a}_{\rm gas}$ only depends on $a$, while  $\dot{a}_\star\propto m^{-0.91}$. Therefore  the stellar hardening rate becomes less efficient as the binary increases its total mass and, overall, $\dot{a}>0$. The binary expansion continues until the binary separation reaches the threshold $a=CR_{\rm sg}$ where the gas-driven expansion is hampered.  The binary semi-major axis then moves along $CR_{\rm sg}$, which decreases as the binary increases its mass ($R_{\rm sg}\propto m^{-7/45}$ for fixed $f_{\rm Edd}$). The accretion continues until the binary mass grows large enough so that $|\dot{a}_\star+\dot{a}_{\rm GW}|>|\dot{a}_{\rm gas}|$ and the binary starts shrinking very efficiently down to its coalescence.

In the scenario where the binary accretes mass at a fixed $\dot{m}$, 
\new{the binary expands only if $\dot{a}_{\rm gas}$ is large enough so that, once $a$ reaches  $C R_{\rm sg}$ from larger separations and gas-driven evolution switches on,  expansion and shrinking find an equilibrium point at  $a=C R_{\rm sg}\propto m^{38/45}$; this dependence on the mass implies that the binary expands owing to its mass growth, since it is bound to move along $C R_{\rm sg}$. The binary expansion proceeds until the binary mass becomes large enough for GWs to take over, ensuring coalescence.}
\new{It is worth stressing that, in the fixed $\dot{m}$ scenario, expansion and significant mass growth only occur for highly super-Eddington initial accretion rates (e.g. in Fig.~\ref{fig:evol_fixed_mdot}, the green curve showing expansion initially assumes  $\dot{m}\sim10^3 \dot{m}_{\rm Edd}$) for an initial binary mass of $10^4M_{\odot}$. } 

\new{The} picture \new{outlined above} clearly implies that, \textit{although the binary may undergo a phase of expansion, this necessarily reverts into shrinking when $m$ becomes large enough, so that the binary will always enter the GW emission phase and reach its final coalescence.}

Given the picture outlined above, we can compute the maximum mass that can be reached by the binary before GWs counteract the effect of gas and ensure a prompt shrinking; note that this mass is an upper limit that may be attained only if there exist a moment in time at which  $\dot{a}>0$.\footnote{This threshold mass, once reached, is very close to the mass at the merger, as the binary shrinks very quickly in the GW regime and is not able to accrete a significant amount of gas.
} We compare $R_{\rm sg}$ with $a_{\rm gas \ \rightarrow \ GW}$ and we obtain, for a fixed $\dot{m}$:
\begin{equation}
\begin{split}\label{eq:mmax_fixmdot}
    \frac{m_{\rm max}}{3.58\times10^8\msun} &= \left(\frac{C}{0.5}\right)^{1.5} \left(\frac{\dot{m}}{\msun\, {\rm yr}^{-1}}\right)^{-0.36}\\
    &\left(\frac{\alpha}{0.1}\right)^{0.93}\left(\frac{4q}{(1+q)^2}\right)^{-0.375};
    \end{split}
\end{equation}
if instead we fix $f_{\rm Edd}$, we find
\begin{equation}
\begin{split}
\label{eq:mmax_fixf}
    \frac{m_{\rm max}}{4.00\times10^7\msun} &= \left(\frac{C}{0.5}\right)^{1.1} \left(\frac{f_{\rm Edd}}{1}\right)^{-0.26}\\ 
    & \left(\frac{\alpha}{0.1}\right)^{0.69}\left(\frac{\eta}{0.1}\right)^{0.54}
    \left(\frac{4q}{(1+q)^2}\right)^{-0.28}.
    \end{split}
\end{equation}
This maximum mass decreases with $f_{\rm Edd}$ or $\dot{m}$, implying that the total mass accreted along the inspiral as a function of  $f_{\rm Edd}$ or $\dot{m}$ peaks near the smallest $f_{\rm Edd}$ or $\dot{m}$ for which $\dot{a}>0$ is attained.\footnote{ The only caveat to this consideration is that the maximum mass has been derived equating $CR_{\rm sg}$ to  $a_{\rm gas \ \rightarrow \ GW}$, in fact assuming that gas-driven expansion is counterbalanced by GW emission only. Stars however may also give a non-negligible contribution to counterbalance the gaseous evolution, so that the maximum  mass can in fact be slightly smaller than the one reported here.} 

\new{Fig.~\ref{fig:evol_fixed_m}  shows the evolution of binaries with different initial mass, assuming a fixed $f_{\rm Edd}=1$. For this choice of the Eddington ratio, all binaries with mass $\lesssim m_{\rm max}$ (evaluated when they cross $CR_{\rm sg}$ the first time) undergo an expansion phase; all their tracks in the $(a, m)$ space overlap during the evolution, so that the final phase is the same. Only the most massive binaries (see brown solid line in Fig. \ref{fig:evol_fixed_m}) do not undergo expansion and may end up with a final mass $\gg m_{\rm max}$. }

\new{It is worth stressing that, in our model, we  assumed the binary to continue growing its mass even for $a>CR_{\rm sg}$. We also performed a series of tests damping mass accretion together with binary expansion at $CR_{\rm sg}$; this can mimic inefficient accretion beyond this radius, or a gaseous disc which is spatially limited to $CR_{\rm sg}$ (or exhausted once this separation is reached). In this situation, we found the overall evolutionary picture depicted here to remain similar and the time needed to reach coalescence to be $\lesssim10^8$ yr, but the specific evolutionary tracks may vary.\footnote{\new{%
If mass accretion is damped together with expansion,  the solution of the differential equations proceeds so that, in case of expansion, the binary stalls at a radius in which expansion is nearly hampered, but still some mass accretion can occur. 
Only if accretion is stopped before the binary reaches  $CR_{\rm sg}$  the binary may endlessly stall owing to the lack of mass growth; however, we believe this latter configuration to be rather unphysical.}}}

\begin{figure}
    \centering
    \includegraphics[width=1\columnwidth]{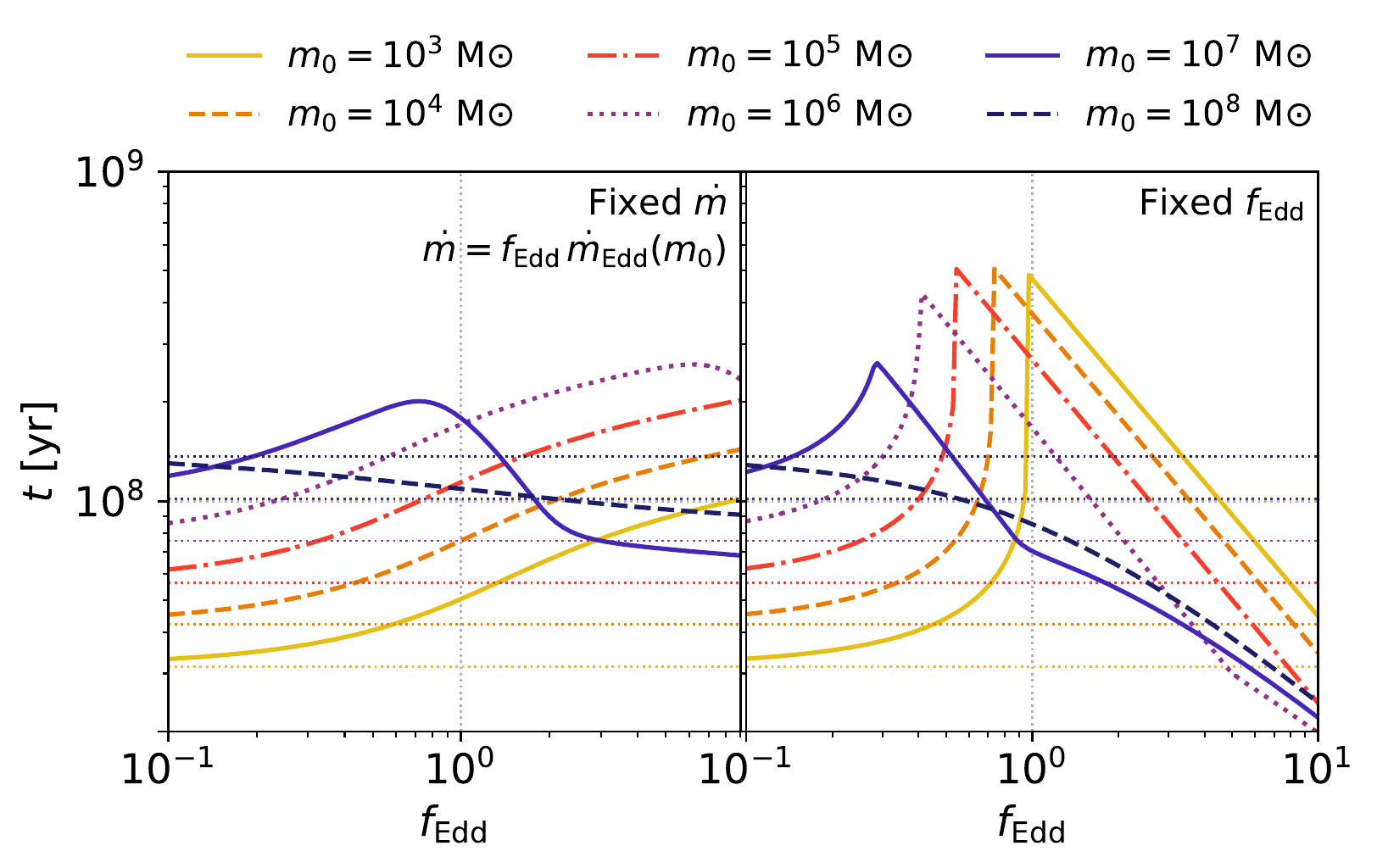}
    \caption{The plot shows the timescale needed by a binary to shrink from $10 a_h$ to the final coalescence, as a function of the Eddington factor, for different binary masses (as coded in the legend). In the left-hand panel, the binary mass accretion $\dot{m}$ is assumed to be fixed in time, and equal to $f_{\rm Edd}$ times the initial Eddington accretion rate ($m_{\rm Edd}$ evaluated at $m_0$). In the right-hand panel we instead assume a fixed Eddington ratio throughout the evolution. The thin dotted horizontal lines mark the inspiral time for non-accreting binaries ($\dot{m}=0$); they refer to different $m_0$, as in the legend. \new{In the right-hand panel, beyond the peak and before the slope change, the inspiral time is well described  via Eq.~\ref{eq:tmax}, as better detailed in the text. }}
    \label{fig:timescales}
\end{figure}

Fig.~\ref{fig:timescales} shows the timescale needed by binaries with different initial mass to reach their final coalescence as a function of the Eddington ratio, for a fixed mass accretion rate (left) or Eddington ratio (right).  Remarkably, \textit{the inspiral timescale does not get dramatically longer as a result of the gas-driven expansion}: if we fix $\dot{m}$ and consider reasonable values  for the mass accretion rate, the inspiral timescale increases by a factor of a few ($<10$) at most. The impact of gas is more relevant when adopting a fixed $f_{\rm Edd}$. In that case, the inspiral time peaks near $f_{\rm Edd}=1$, and specifically very close to the smallest $f_{\rm Edd}$ allowing for $\dot{a}>0$ to be attained along the evolution. The inspiral time  declines for large $f_{\rm Edd}$ as the binary efficient mass growth implies that all processes occur faster. Also note that, in proportion, the inspiral time is less impacted by gas-driven expansion as the binary initial mass is larger: when the initial mass is closer to the limiting mass in Eq.~\ref{eq:mmax_fixmdot}~or~\ref{eq:mmax_fixf}, the binary needs a shorter time to get there and enter the GW-driven inspiral.

Clearly, the specific evolutionary timescales obtained here depend both on the choice of $C$, i.e. the multiple of $R_{\rm sg}$ at which gas expansion is damped, and on how sharp is the suppression at such point. Nevertheless, we checked that the global scenario depicted here is robust against variations of such quantities.

\new{Finally, we can also estimate the inspiral time of binaries if expansion ever occurs. A fixed  $f_{\rm Edd }$ implies that  $dm/m = f_{\rm Edd } \left[4\pi G \, m_{\rm p} /(\eta \sigma_{\rm T} c) \right] dt $, so that the mass grows exponentially, and the time needed to reach $m_{\rm max}$ from $m_0$ (computed via Eq.~\ref{eq:mmax_fixf}) naturally follows from the Salpeter timescale:}
\begin{equation}
\begin{split}\label{eq:tmax}
    t_{\rm exp} = \log\left(\frac{m_{\rm max}}{m_0} \right) \,  \frac{1}{f_{\rm Edd}}\, \frac{\eta \sigma_{\rm T} c}{4\pi G \, m_{\rm p}} = \\ 4.415 \times 10^7\, \left(\frac{f_{\rm Edd}}{1}\right)^{-1}\log\left(\frac{m_{\rm max}}{m_0} \right)\, {\rm yr}.
\end{split}    
\end{equation}
\new{The value of $t_{\rm exp}$ is in very good agreement with our results, and nearly perfectly overlaps with the curves in the right-hand panel of Fig.~\ref{fig:timescales} beyond the peak (except for values of $f_{\rm Edd}$ so large that  $m>m_{\rm max}$ as $CR_{\rm sg}$ is first reached). In principle, one could make an analogous estimate for the cases with fixed $\dot{m}$, but this would be of little value as expansion is almost never attained if $\dot{m}$ is fixed.}

%%%%%%%%%%%%%%%%%%%%%%%%%%%%%%%%%%%%%%%%%%%%%%%%%%

%%%%%%%%%%%%%%%%%%%%%%%%%%%%%%%%%%%%%%%%%%%%%%%%%%

\section{Discussion and conclusions}\label{sec:discussion}

In this study we addressed the evolution of MBHBs under the concurrent effect of different mechanisms, specifically the gas-, stellar- and GW-driven evolution. In particular, we explored the consequences of an outspiral phase mediated by the MBHB interaction with a gaseous circumbinary disk, based on the results of recent numerical simulations \citep{Tiede2020, Munoz2020, Duffell2020}. We described the MBHB evolution through a coupled set of simplified differential equations expressing the time variation of the semi-major axis and the total mass. Specifically, stars, gas and GWs all contribute to the semi-major axis evolution, while the mass ratio and total mass growth depend on gas dynamics only.

The key result of the present paper is that  a putative phase of gas-driven expansion does not sensibly impact the MBHB evolution and inspiral time. Binary expansion can occur only if the binary mass accretion rate remains close or exceeds the Eddington limit along the entire evolution. In these cases, the binary may undergo an outspiral phase. However,  it cannot expand  to scales larger than the self-gravitating radius (i.e. the radius at which disk self-gravity cannot be further ignored, and gas-driven expansion can no longer occur; \citealt{franchini2021}).  Following this possible expansion phase, the semi-major axis nearly stalls at the self-gravitating radius, while the binary continues accreting, until its mass gets large enough for GWs to induce coalescence. Irrespective of the presence of an outspiral phase, we always find that MBHBs reach the coalescence within at most few hundreds of Myr. Thus, although binaries interacting with gas can undergo an expansion phase, our findings suggest that such phase does not hinder MBHBs  coalescence. \new{Noticeably, a putative expansion would lengthen the phase at which the binary and the disc are coupled at sub-pc scales (see e.g. Fig.~\ref{fig:evol_fixed_m}). Therefore, according to the presented scenario, the luminous sub-pc binaries targeted by present and future time-domain surveys  could be more common than previously thought.}

Given the simplifications made in the current implementation, our  treatment is subjected to a number of caveats that we now discuss. We would like however to stress that all our assumptions were \new{in general} conservative.

We assumed, for simplicity, that once the binary enters the mass regime where the disk self-gravity is important, the gas-driven expansion  is halted. The interaction of a binary with a self-gravitating accretion disk has been studied through hydrodynamical simulations by e.g. \citet{Cuadra2009,Roedig2012,franchini2021}. These works focused on self-regulated disks, i.e. for which the cooling is not efficient enough for the disk to fragment into bound clumps, and found that the interaction with the binary leads to its shrinking in the wide region of the parameters space explored. If the disk instead does fragment into clumps, possibly forming stars, the interaction with the binary would occur by means of stellar slingshot ejections, which induce the shrinking (see Sec.~\ref{sec:stellar_hardening}). Therefore the binary semi-major axis decreases with time in both scenarios, making our choice to suppress the evolution quite conservative. 
Furthermore, if the structure of the circumbinary disk surrounding a MBHB is similar to the  structure  of  accretion  disks in AGN \citep{CollingSuffrin1990}, then the disk aspect ratio is expected to be $H/R \approx 0.01-10^{-3}$. In this regime of disk aspect ratios, the binary is found to shrink due to the interaction with the disk even in the non-self-gravitating regime.
Moreover, it has been shown that binaries with \new{initial values for $q<0.05$ or $e>0.1$ tend to a solution that guarantees the inspiral,} regardless of the exact properties of the circumbinary disk, \new{suggesting that even a mild initial eccentricity could avert the expansion} \citep{Dorazio2021}.

Our modelisation of the stellar hardening (Sec.~\ref{sec:stellar_hardening}, Eq.~\ref{eq:stellar_hardening}), was based on well established scaling relations linking the density and velocity dispersion with the MBHB total mass \citep[see e.g.][]{Sesana2015}. This provides an ``average'' connection between MBHs and the characteristics of their hosts. This is of course a simplification, but it is conservative in many ways. First of all, the employed scaling relations do not consider the possible presence of a nuclear star cluster in the galaxy centre, which would enhance the hardening by increasing the density at the binary influence radius. Therefore the evolutionary timescale   found with our implementation should be considered as a conservative estimate of the stellar-driven shrinking, which in many galaxies may be much more efficient. In addition we updated $\rho$, $\sigma$ that enter the stellar hardening prescription (Eq.~\ref{eq:stellar_hardening}) with the binary total mass at any given time; this implies a symbiotic evolution of the binary and the galaxy, although the galaxy properties are likely to evolve on a much longer timescale compared with the binary evolution. Note that this assumption is also conservative, as $\dot{a}_\star \propto \rho/\sigma \propto m^{-0.91}$, implying that we would get a more efficient stellar hardening if we were to keep fixed the galaxy properties or to make them evolve more slowly as the binary mass grows. In particular, this assumption is what drives the expansion phase when $f_{\rm Edd}$ is fixed, as described above, and a less conservative assumption on $\dot{a}_\star$ may result in binary stalling phase rather than an expansion phase (which would anyways end with coalescence due to the relentless mass accretion). The  important caveat here is that the stellar content of galaxies in the very high-$z$ universe may have been relatively low, so that the employed scaling relations may not be adapt for the early Universe, in which much of the gas content still had to be turned into stars. However, the structural parameters of high-$z$ galaxies near their centre, their gas and stellar content and morphological properties are poorly known, and the investigation of this aspect is beyond the scope of the present work.

Another assumption we made in this work concerns the binary eccentricity, which we force to remain zero throughout the binary evolution.
This is again a conservative assumption, as the binary eccentricity may be non-zero at the binary formation time and may  grow during the stellar-hardening  phase, especially if the stellar background is isotropic or counter-rotates with the binary.
Furthermore, the interaction of an initially circular or slightly eccentric binary with a gaseous circumbinary disk is also found to impact the binary eccentricity  in both the non-self-gravitating \citep{Ragusa2020} and the self-gravitating case \citep{Roedig2011}.
A non-zero eccentricity can impact both stellar and GW-driven hardening. The effect on the former is generally a mild enhancement of the hardening efficiency by stars, while for the latter it provides a strong boost in the emitted GW power which strongly increase as binaries become eccentric, determining a GW-driven inspiral to become much more efficient compared with the circular cases.

Finally, the existence of a binary expansion phase implies a significant mass growth prior to merger. Although this might shift a significant number of low mass MBHBs at the margin of the LISA sensitivity band (LISA will be mostly sensitive to MBHBs of $M<10^7\msun$), it is important to bear in mind that we assumed an unlimited supply of gas. In reality, most of the cold gas funneled to the center of a merger remnant is converted in stars and a circumbinary disk cannot be sustained indefinitely. A putative stalling or expansion phase may in practice cut off by the consumption of the available cold gas reservoir. To fully assess the putative effect of binary outspiral on the population of MBHBs observable by LISA and PTA and on the statistics of MBHBs possibly observable as electromagnetic periodic sources, our semianalytic model should be coupled to a framework describing the evolution of MBHs and their hosts along the cosmic history \citep{2020MNRAS.495.4681I}, \new{in analogy with what was done by e.g. \citet{Kelley2017b} (who however did not account for gas-driven expansion) in the context of cosmological simulations}. This is deferred to future work.

\begin{acknowledgments}

We warmly thank the anonymous referee for their useful comments and suggestions, and Massimo Dotti for fruitful discussion.
EB, AF and AS acknowledge financial support provided under the European Union’s H2020 ERC Consolidator Grant "Binary Massive Black Hole Astrophysics" (B Massive, Grant Agreement: 818691). MB acknowledge funding from MIUR under the grant PRIN 2017-MB8AEZ. 
\end{acknowledgments}

\bibliography{bibliography}{}
\bibliographystyle{aasjournal}

\end{document}